# Infrared spectra of solid indene pure and in water ice. Implications for observed IR absorptions in TMC-1.

Belén Maté[1*], Isabel Tanarro[1], Vicente Timón[1], José Cernicharo[2] and Victor J. Herrero[1]

1. Instituto de Estructura de la Materia, IEM-CSIC, Serrano 121-123, 28006 Madrid, Spain.

2. Instituto de Física Fundamental, IFF-CSIC, Serrano 123, 28006, Madrid, Spain

*Corresponding author: Belén Maté.

Email: belen.mate@csic.es

**Abstract**

Experimental and theoretical infrared spectra, between 4000-500 cm$^{-1}$ (2.5-20 μm), and infrared band strengths of two solid phases of indene, amorphous and crystalline, are given for the first time. The samples were generated via vapor deposition under high vacuum conditions on a cold surface. Density functional theory was employed for the calculations of the IR spectra. Lacking of previous information, a monoclinic symmetry is suggested for the theoretical crystalline phase of indene, based on the comparison of the calculated and experimental IR spectra. Assignments, based on the calculations, are given for the main indene IR absorptions. The infrared spectra of highly diluted mixtures of indene in amorphous solid water at 10 K are also provided, evidencing that the indene spectrum is not much altered by the water ice environment. These data are expected to be useful for the search of this species in the solid phase in astrophysical environments with the JWST.

With the band strengths obtained in this work, and applying a simple literature model, we find that indene could represent at most 2-5% of the intensity of a weak absorption feature at 3.3 μm recently reported for Elias 16. A column density of (1.5 -0.6) × 10$^{16}$ cm$^{-2}$ is estimated for indene in the ice mantles of TMC-1. It would correspond to ≈ (2 - 0.8) × 10$^{-2}$ of cosmic carbon, which is probably too high for a single small hydrocarbon.

## 1. Introduction

In the mid-eighties, Léger et Puget (1984) and Allamandola et al. (1985) proposed that based on the IR fluorescence properties of laboratory polycyclic aromatic hydrocarbons (PAHs), some related species were the likely carriers of the unidentified infrared emission (UIE) features (Willner et al. 1979), found in interstellar regions illuminated by UV photons, leading to the so called PAH hypothesis. The PAH hypothesis (Tielens 2008) has been widely, but not universally, accepted and a long-standing criticism (Zhang and Kwok 2015) was the non-detection of individual PAH molecules in the interstellar medium (ISM). The detection of individual pure PAHs is far from trivial. The identification of individual molecules in the ISM is mostly achieved through radiofrequency measurements of specific rotational transitions, and these measurements are very difficult for PAHs because they usually have negligible or very small dipole moments coupled to low fractional individual abundances. In 2021, the first PAH molecule, indene, was finally detected in the cold pre-stellar core TMC-1 (Cernicharo et al. 2021, Bukhardt et al. et al. 2021). Indene is a bicyclic molecule composed of a six- and a five-membered ring and has a small but appreciable (≈ 0.6-0.7 D) dipole moment (Caminati 1993, Burkhardt et al. 2021). Relatively high abundances, for such a complex molecule, of gas-phase indene (1-1.6 × $10^{-9}$ with respect to $H_2$) were derived from the observations.

It has been traditionally assumed that PAHs should be largely formed in the envelopes of carbon rich asymptotic giant branch (AGB) stars, starting from simple molecular species like $C_2H_2$ (Frenklach & Feigelson, 1989; Cherchneff, 2012), or at later stages of stellar evolution (Kwok 2004), like protoplanetary nebulae (PPN), where benzene has been detected (Cernicharo et al. 2001, Malek et al. 2011). In these environments, high temperatures, densities, or UV fields favor a chemistry conducting to the formation of aromatic structures (Woods et al. 2003, Cernicharo 2004, Martínez et al. 2020, Santoro et al. 2020). However, the first detection of a specific gas-phase pure PAH molecule has taken place in a cold cloud. Besides indene ($C_9H_8$), the cyano derivatives of benzene ($C_6H_5CN$), indene ($C_9H_7CN$), and naphthalene ($C_{10}H_7CN$) were also found in TMC-1 (McGuire 2018, McGuire 2021) with estimated abundances in the $10^{-10}$-$10^{-11}$ range with respect to $H_2$. The nitrile groups in these molecules largely increase their dipole moments with respect to their pure hydrocarbon counterparts and makes them thus much easier to detect. It has been suggested that they could be taken as proxies for the non-functionalized molecules. Estimates based on the observed $C_9H_8$/$C_9H_7CN$ quotient and on a chemical model, indicate that the pure hydrocarbons should be ≈ 20-40 times more abundant than their cyano derivatives (Sita et al. 2022). In this context it is worth noting that benzyne (c-$C_6H_4$) and cyclopentadiene (c-$C_5H_6$) have been also found in the same cold dark cloud (Cernicharo et al., 2021a, 2021b).

Volatile molecules like indene should freeze to a large extent on the ice mantles of dust grains in TMC-1. In fact, the presence of PAHs in the ice mantles of cold clouds has been conjectured since the start of the PAH hypothesis and possible signatures of PAHs in the IR ice absorption spectra have been sought. In the nineties, a weak band at about 3.25 μm, detected in some clouds, was tentatively attributed to the CH stretch vibration of polyaromatic molecules (Sellgren 1995, Brooke 1999), and in a recent work, Chiar et al. (2021) have suggested that CH stretching vibrations of $CH_2$ groups in hydrogenated PAHs could account in part for the widely observed feature at 3.47 μm which is mostly attributed to ammonia hydrates (Dartois & D' Hendecourt 2001, Dartois et al. 2002, Boogert el al. 2015). It was also surmised that other characteristic bands of PAHs like the CC stretching vibrations at 6.2 μm (Keane et al. 2001) and the CH out of plane bending vibrations at 11, 2 μm (Bregman 2000) could also contribute somewhat to other ice absorption features. However, the suggested contribution of PAHs to specific ice absorption bands remains highly speculative. The mentioned IR features appear as very weak substructures of broader absorptions and are much affected by uncertainties in baseline subtraction, and by telluric absorptions in ground-based observations. In addition, some of these substructures can have a contribution from ice profile modifications or from other ice species (Boogert et al. 2015). Using laboratory data and theoretical calculations for the IR band strengths of representative polyaromatic species, the maximum amount of cosmic carbon locked up in PAHs compatible with an attribution the absorption features in the ices of dense clouds was estimated at ≈ 5-15% (Bowman et al. 2011, Hardegree-Ullman et al. 2014, Chiar et al. 2021), which is consistent with the abundances derived for PAHs in photon dominated regions of the ISM (Tielens 2008) where the aromatic infrared emission bands are seen. In general, it was assumed that the hypothetic polyaromatic hydrocarbons in the ice mantles of dense clouds should be mixtures of large PAHs (≈ 50 C atoms or more), able to withstand the intense UV fields in the transit from post AGB regions to dark clouds. The abundant presence of small gas-phase PAHs like indene and, presumably, naphthalene inside the cold dense cloud TMC-1 (Cernicharo et al. 2021, Burkhardt et al. 2021) is thus puzzling, since they are too small to survive the UV field outside the clouds, which strongly suggests that these molecules have been formed in situ. However, current models, including gas-phase and grain surface reactions, underpredict the observed indene abundances in TMC-1 by three orders of magnitude (Doddipatla et al. 2021, Burkhardt et al. 2021). The high relative abundance of gas-phase indene in TMC-1 raises the question about the cycling of this molecule between the gas and the solid. In the interior of dense clouds, in the absence of significant thermal of photo desorption, cosmic ray (CR) sputtering could provide a way to balance the freezing and accretion of indene on solid grains. This possibility has been recently addressed by Dartois et al. (2022) who used laboratory data on ion sputtering of PAHs to estimate the gas-solid partition of naphthalene in a dense cloud. They concluded that the CR sputtering mechanism would need fractional abundances of $10^{-3}$-$10^{-4}$ of naphthalene molecules in the ice mantles to justify the gas-phase abundances estimated for TMC-1.

Progress in the understanding of the high abundance of indene in TMC-1 requires further modeling and laboratory data. Experimental IR spectra of indene are available in the literature for the liquid and for the vapor (Klots 1995), and theoretical calculations, showing good agreement with the measurements, have been published for the isolated molecule (Klots 1995, El-Azhary 1999), but as far as we know no IR spectra of indene ices have been reported. The present work is focused on the IR spectroscopy of solid phases of indene at low temperatures. IR spectra of vapor deposited amorphous and crystalline indene and of indene mixtures with water ice have been recorded. Solid structures and vibrational spectra have been calculated using density functional theory (DFT) and the results of the calculations have been used for the assignment of the measured IR spectra. Experimental and theoretical band strengths have also been determined and the astrophysical implications of the results for the IR spectra of ices in TMC-1 are discussed.

## 2. Experimental section

The experimental set-up has been described previously (Maté et al. 2021). It consists in a high-vacuum chamber, with a background pressure in the $10^{-8}$ mbar range, provided with a closed-cycle He cryostat and coupled to a FTIR spectrometer (Bruker Vertex70) through KBr windows. Indene and water vapors are introduced in the chamber through independent lines with needle valves, and condensed on a cold Si substrate placed in thermal contact with the cold head of the He cryostat. Indene is a commercial liquid (≥99%, Sigma-Aldrich) with about 1 mbar vapor pressure at 20°C. In our experimental setup, due to the small fluxes provided by the needle valves, it is convenient to increase the indene vapor pressure, by immersing a Pyrex tube containing the liquid in a silicone oil bath at 60°C. In this way the indene pressure s increased up to about 14 mbar. To guarantee the stability of water vapor pressure during the deposit, a water flask is placed in a thermal bath at 30°C. Both lines are heated to avoid condensation. Pure indene layers were generated at 10 K and at 160 K, with a deposition vapor pressure in the $10^{-6}$ mbar range. Ice mixtures were generated at 10 K with vapor pressure in the $10^{-5}$ mbar range. Ice layers of thickness of several tens of nm were grown.

A quadrupole mass spectrometer (Hiden200) (QMS) directly connected to the HV chamber allows monitoring gas phase in the chamber during the experiments. A Faraday cup was used as detector. Absolute gas-phase densities in the deposition chamber were estimated by calibrating the quadrupole mass spectrometer as described in Appendix A1.

Normal incidence transmission spectra with a 4 $cm^{-1}$ resolution and 100 scans accumulation were recorded with the Vertex 70 FTIR spectrometer and a liquid nitrogen cooled mercury cadmium telluride (MCT) detector.

## 3. Theoretical calculations

The structure of solid crystalline indene is not known from bibliographic data. In this work, in a first attempt, it was guessed the indene crystal symmetry to be orthorhombic, in analogy to that of the crystal of i3-(phenylthio)indene, with molecular formula $C_{15}H_{12}S$ (Curnow et al. (2012)). To build the initial unit cell the structure of $C_{15}H_{12}S$ molecular crystal was taken, transforming the molecular unit $C_{15}H_{12}S$ to $C_8H_9$ molecule by removing the phenyl group and the S atom. However, the calculated IR spectra of the relaxed solid obtained in this way do not show a clear resemblance with the experimental spectra of crystalline indene. In a second attempt, a monoclinic symmetry was postulated. This idea was inspired from the behavior of the molecular crystal of 1-Bromo-2,3,5,6-tetramethylbenzene, also known as bromodurene, molecular formula $C_{10}H_{13}Br$, which, depending on temperature, crystallize in an orthorhombic or in a monoclinic system (Hamdouni et al 2019). The IR theoretical spectra obtained in this case is in good agreement with the experimental one, and this was taken as proof of the goodness of the hypothesis.

Geometry optimization was carried out with the CASTEP code (Clark et al 2005) using the Broyden–Fletcher–Goldfarb–Shanno optimization scheme (Payne et al. 1992) under density functional theory (DFT)-based methodology. The exchange-correlation energy term was treated using the generalized gradient approximation (GGA) with revised Perdew-Berke-Ernzerhof (rPBE) settings (Zhang et al. 1998). Since indene molecules are packed together inside the unit cell, probably via van der Waals forces and/or weak and moderate hydrogen bonds, to better consider those interactions the Tkatchenko–Scheffler (TS) dispersion correction was included in the calculations (Tkachenko et al. 2009). OTFG pseudopotentials (Pickard et al. 2000) with a cut-off 720 eV were employed. The convergence criteria were set at $1 \times 10^{-5}$ eV/atom for the energy, 0.05 eV/Å for the interatomic forces, maximum stress 0.1 GPa and 0.002 Å for the displacements. Atomic forces and charges were evaluated at the minimum in the potential energy surface to predict the harmonic vibrational infrared spectrum by means of density functional perturbation theory (Baroni et al. 2006).

On the other hand, no information about the structure of amorphous indene is available, in particular, on its density. In order to simulate an amorphous indene solid, a unit cell was constructed containing four molecules with an initial density of 1.3 g/cm$^3$, chosen to match that of the liquid phase of indene. Amorphous indene structures were generated through fast force-field molecular dynamics (MD) simulations using the Amorphous Cell modulus of the Materials Studio (MS) package. The classical MD simulations implemented were then optimized by means of periodic quantum mechanical calculations, allowing the system to modifying not only the molecular arrangement but also the unit cell size, and therefore its density. Its IR spectra were calculated with the same methodology described for the crystal.

## 4. Results and discussion.

### 4.1. Amorphous and crystalline indene and phase transition.

Figure 1 shows the mid IR spectra of amorphous and crystalline indene layers grown by vapor deposition at 10 K and at 160 K, respectively. Important differences in the IR spectra of the two phases are evidenced in the CH stretching region around 3000 cm$^{-1}$ (top panel), where frequency shifts are appreciable and the relative intensities of the peaks change considerably. At lower wavenumbers there are also substantial intensity variations upon crystallization, but no significant band shape changes. A list of the main IR peaks of indene is given in Table 1, both for amorphous and crystalline phases. The peak positions of some minor features discernible in the spectra are not given in the table. An assignment of the strongest bands to vibrational modes of the indene molecule has been made, guided by the calculations performed in this work and presented in section 4. The mode assignments will be discussed in that section. Although there are works that present the IR spectra of the indene molecule and the liquid, providing IR band assignment (Klots1995), as far as we know, the IR spectra of the solid phases of indene are presented here for the first time.

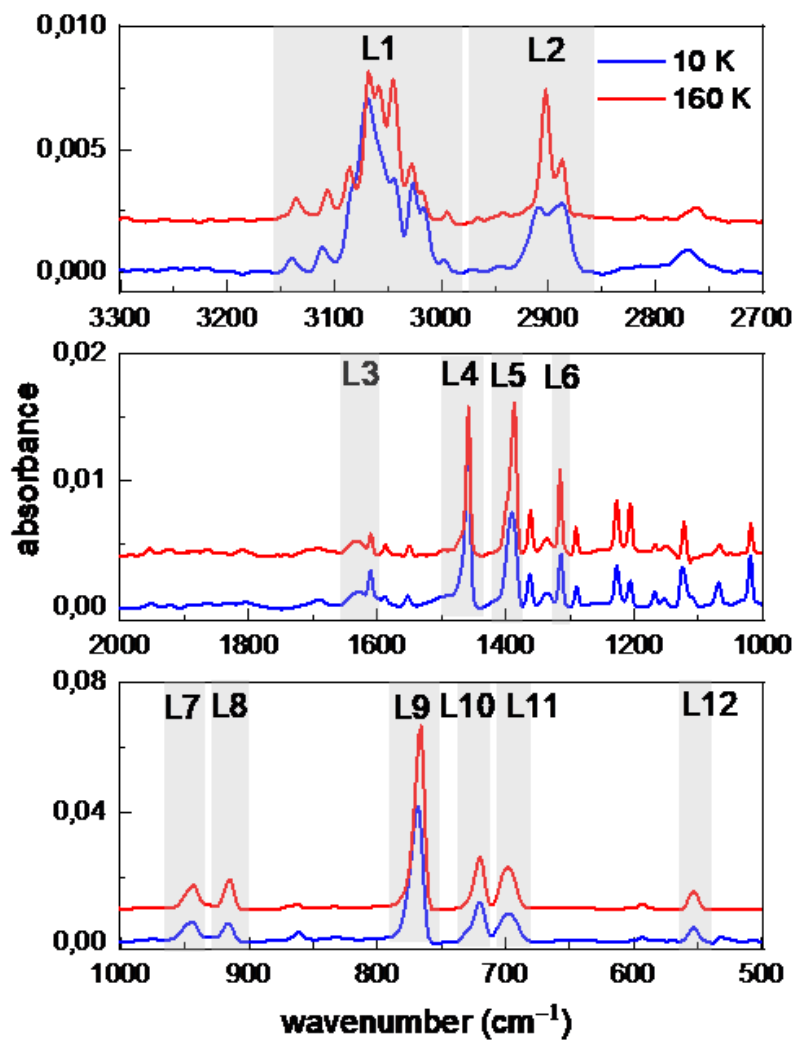

Figure 1. Infrared spectra of amorphous (blue) and crystalline (red) indene layers grown by vapor deposition at 10 K and 160 K, respectively. The estimated thickness, assuming a density of 1.3 g/cm3, is 20 nm per each side of the Si substrate.

Table 1. Peak positions of the main absorptions of amorphous and crystalline indene solids at 10 K and 160 K, respectively. The mode assignment is based on the theoretical calculations performed in this work, presented in section 4.3. When the wavenumber corresponds to a shoulder of a stronger band it is indicated as "sh", and when the peak corresponds to a broad band it has been indicated with "br". $\nu_a$ indicate asymmetric stretching modes, $\delta$ bending modes.

| Peak position (cm$^{-1}$) | Assignment |
| --- | --- |

| amorphous | crystalline | lavel | mode |
|---|---|---|---|
| 3139,3111,3084 | 3139, 3106,3086 | L1 | $\nu_a$ CH |
| 3069 | 3069 | | CH hexa- ring + CH penta- ring |
| 3044sh,3026,3016 | 3044sh,3028,3016 | | |
| 2909, 2891,2887 | 2909, 2886 | L2 | $\nu_a$ CH$_2$ |
| 2770 | 2771 | | |
| 1688 | 1708, 1690 | | |
| 1633 br | 1637, 1630 | L3 | $\nu$ C=C |
| 1610 | 1609 | | |
| 1587 | 1587 | | |
| 1552 | 1549 | | |
| 1470 sh | 1471 sh | | |
| 1459 | 1458 | L4 | $\delta$ in-phase in-plane 3 CH hexa-ring |
| 1390 | 1399 sh | L5 | $\delta$ CH2 |
| | 1387 | | |
| 1362 | 1361 | | |
| 1336 br | 1336 | | |
| 1314 | 1315 | L6 | $\delta$ in-phase in-plane 2 CH penta-ring |
| 1289 | 1290 | | |
| 1227 | 1227 | | |
| 1206 | 1206 | | |
| 1168 | 1168 | | |
| 1153 | 1151 | | |
| 1125 | 1122 | | |
| 1108 sh | 1106 sh | | |
| 1068 | 1067 | | |
| 1019 | 1018 | | |
| 977 w,br | 982 w, br | | |
| 945 | 944 | L7 | $\delta$ out-of-phase in-plane 4 CH hexa-ring |
| 916 | 915 | L8 | $\delta$ in-plane CH-CH$_2$(block) |
| 862 | 870, 863 | | |
| 832 | 833 | | |
| 768 | 766 | L9 | $\delta$ in-phase out-of-plane 8CH |
| 730 sh | 730 sh | | |
| 720 | 720 | L10 | $\delta$ in-phase out-of-plane 4 CH hexa-ring |
| 698 | 698 | L11 | $\delta$ in-phase out-of-plane 2 CH penta-ring |
| 593 | 593 | | |
| 553 | 553 | L12 | Out-of-plane deformation hexa and penta rings |
| 532 | | | |

When a pure indene ice grown at 10 K was heated with a continuous ramp of 1 K/min, the phase transition was observed to start at 120 K and to end at 130 K. Indene sublimation started at 180 K, well above the water sublimation temperature. Figure 2 shows the spectra at different temperatures during the annealing process of indene.

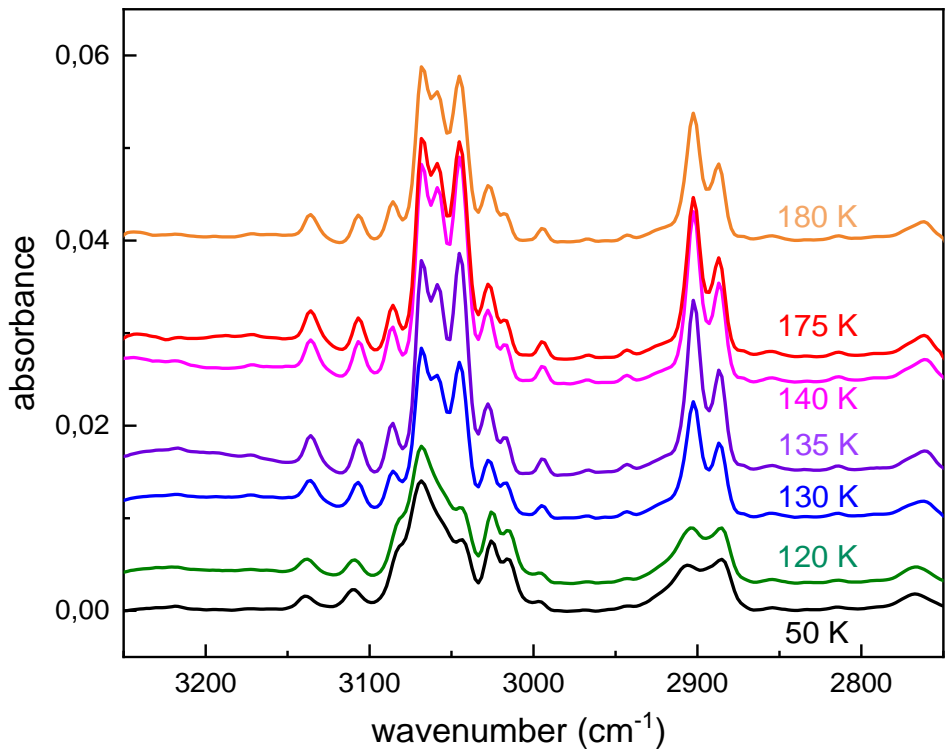

Figure 2. Enlargement of the CH stretching region of the IR spectra of indene, showing the evolution with temperature of an indene layer grown at 10 K and warmed at 1 K/min. Spectra are shifted in the absorbance axis for a better visualization.

**4.2. Infrared band strengths**.

The infrared band strength of a given band (band i) of solid indene is experimentally obtained via the following expression:

$$A'_i = \frac{\int_i \tau d\tilde{v}}{N} = \frac{2.303 \int_i Abs(\tilde{v}) d\tilde{v}}{N} = 2.303 \frac{Int_i}{N}$$

(1)

where τ is the optical depth, $\tilde{v}$ the wavenumber frequency, Abs the absorbance spectrum, $Int_i$ the integrated area of band *i* in the absorbance spectrum, and $N_{indene}$ the column density of indene. The column density of indene molecules has been estimated via the kinetic theory of gases, from the impinging rate of indene molecules on the cold surface and assuming a sticking probability of one at 10 K:

$$N_{Indene} = n_{Indene} \left(\frac{kT}{2\pi m}\right)^{1/2} \Delta t$$

(2)

where $n_{Indene}$ is the molecular density of gas-phase indene in the chamber, which was kept constant during deposition, $k$ is the Boltzmann constant, $T$ the gas temperature, $m$ the mass of the indene molecule, and $\Delta t$ the deposition time. Absolute $n_{Indene}$ during deposition was estimated by calibrating the quadrupole mass spectrometer with a procedure described in Appendix A1, and already employed in a previous work (Maté et al. 2017). The pressure homogeneity in the chamber assumed within this calibration procedure has been proved in a previous publication by our group (Maté et al 2003), where an agreement better than 10% was found in the ice layer thickness measured via kinetic theory of gases and laser interferometry. In the present work, ice layers of different thickness, varying from 10 ML to 600 ML (1ML= $10^{15}$ molec/cm$^2$), were employed for the estimation of the band strengths. Errors for these magnitudes were estimated to be about 30%, mainly due to the uncertainty in the number density determination, derived by errors in determining the calibration factor of the QMS.

Table 2. Experimental peak positions and band strengths of the main IR absorption bands of amorphous and crystalline indene. The limits taken to calculate the band area in the absorbance spectra are given in columns three and six. Uncertainties in the band strengths are about 30%, due to the uncertainty in number density determinations. The last column shows the intensity variations of the different modes upon crystallization.

| Band label | Peak position amorph. (cm$^{-1}$) | Integration Limits (cm$^{-1}$) | A$_{amorphous}$ (x10$^{-18}$ cm/molec) | Peak position cryst. (cm$^{-1}$) | Integration Limits (cm$^{-1}$) | A$_{cryst}$ (x10$^{-18}$ cm/molec) | A$_{cryst}$/A$_{amorf}$ |
|---|---|---|---|---|---|---|---|
| L1 | 3100 | 3153-2983 | 13,6 | 3100 | 3157-2985 | 10,5 | 0,77 |
| L2 | 2900 | 2920-2873 | 4,2 | 2900 | 2926-2877 | 3,1 | 0,74 |
| L3 | 1620 | 1660-1600 | 1,5 | 1620 | 1660-1600 | 1,4 | 0,96 |
| L4 | 1458 | 1487-1443 | 6,4 | 1458 | 1439-1483 | 4,8 | 0,75 |
| L5 | 1390 | 1419-1374 | 5,1 | 1387 | 1414-1371 | 6,8 | 1,34 |
| L6 | 1314 | 1321-1302 | 1,4 | 1315 | 1327-1304 | 2,2 | 1,54 |
| L7 | 945 | 966-928 | 3,8 | 943 | 966-930 | 4,6 | 1,22 |
| L8 | 916 | 930-904 | 2,6 | 916 | 926-903 | 3,9 | 1,51 |
| L9 | 768 | 793-755 | 20,0 | 766 | 796-752 | 23,0 | 1,15 |
| L10 | 720 | 740-711 | 6,0 | 720 | 741-712 | 7,1 | 1,19 |
| L11 | 698 | 711-678 | 6,1 | 698 | 712-681 | 7,7 | 1,26 |
| L12 | 553 | 565-546 | 1,8 | 553 | 567-538 | 2,5 | 1,42 |

The infrared band strengths of the strongest bands in the IR spectra of indene are listed in Table 2. They have been labeled L1-L12 as in Table 1, where the assignment of the different bands is given. The last column in Table 2 shows the intensity variations of the different modes upon crystallization. It can be noticed that the asymmetric stretching CH modes in the 3000 cm$^{-1}$ region decrease their intensity while the intensity of the low wavenumber modes associated mainly to CH bending vibrations follows the opposite behavior.

It is illustrative to compare the band strength provided in this work for indene with those recently published for ices of other aromatic species, benzene and pyridine by Hudson et al. 2022. Here, the authors measure the thickness and density of benzene or pyridine ices and use those magnitudes to determine the column density needed to extract infrared band strengths form the IR spectra. Band strengths of 1.62 10$^{-17}$ cm/molec and 1.55 10$^{-17}$ cm/molec were estimated for the strongest benzene IR absorption band, for the amorphous and crystalline forms, respectively. The band appears at 676 cm$^{-1}$ in the amorphous form (10 K) ant it splits in a doublet at 679 and 681 cm$^{-1}$ in the crystal (100 K). The IR spectrum of solid pyridine presents a strong absorption at 705 cm-1 (amorphous form) or at 712 cm$^{-1}$ (crystalline form), with an intensity of 0.88 10$^{-17}$ cm/molec (amorphous form) or 1.15 10$^{-17}$ cm/molec (crystalline form). Although Hudson et al. (2022) do not provide a spectral assignment for these bands, it seems reasonable to assign them to the same mode than the strongest band of indene, that appears at 678 cm$^{-1}$. This mode corresponds to the in-phase out-of-plane CH bending and it has a band strength of 2 10$^{-17}$ cm$^{-1}$/molec for amorphous indene (10 K) and 2.3 10$^{-17}$ cm$^{-1}$/ molec for the crystalline form (see Table 2). Therefore, since the number of CH groups per molecule are different for pyridine, benzene or indene, being 5, 6 and 8 respectively, a correlation is observed between the number of CH groups involved in the vibration and the band strength. The larger the number of CH groups involved, the larger the band strength of the vibration. This tendency somehow can be taken as a test of the quality of the band strengths given. IR absorption band strengths were also provided by Sandford et al. 2004 for amorphous naphthalene at 15 K. These authors used water ice bands for the calibration of absolute band intensities in naphthalene mixtures with water. Their band strengths are comparable to those commented on in this paragraph. Specifically, they derived a value of 1.2 × 10$^{-17}$ cm/molec for the out of plane bending vibration at 784.4 cm$^{-1}$. This value is smaller than that from the present work, but the two values lie within their mutual experimental uncertainty.

### 4.3 Theoretical model. Comparison with experiments.

As described in the theoretical section, a monoclinic symmetry was assumed for the indene crystal, containing two indene molecules in the unit cell. The relaxed structure obtained after running the calculations has a density of 1.40 g/cm$^3$. On the other hand, for amorphous indene, a unit cell containing four indene molecules has been constructed, and the relaxed structure reached a density of 1.27 g/cm$^3$. The cell parameters of the optimized structures calculated in this work for crystalline and amorphous indene solids are presented in Table 4, and the structures are shown in Figure 3. More details are given in the Suplementary Material.

Table 4. Cell parameters for calculated amorphous and crystalline indene solids.

| Structure | a | b | C | α | β | γ | Volumen Å³ | Dens(gm/cm³) |
|---|---|---|---|---|---|---|---|---|
| amorphous | 10.12 | 8.48 | 7.49 | 98.69 | 72.25 | 94.99 | 605.12 | 1.27 |
| crystalline | 4.98 | 4.87 | 12.34 | 90.00 | 67.05 | 90.00 | 276.11 | 1.40 |

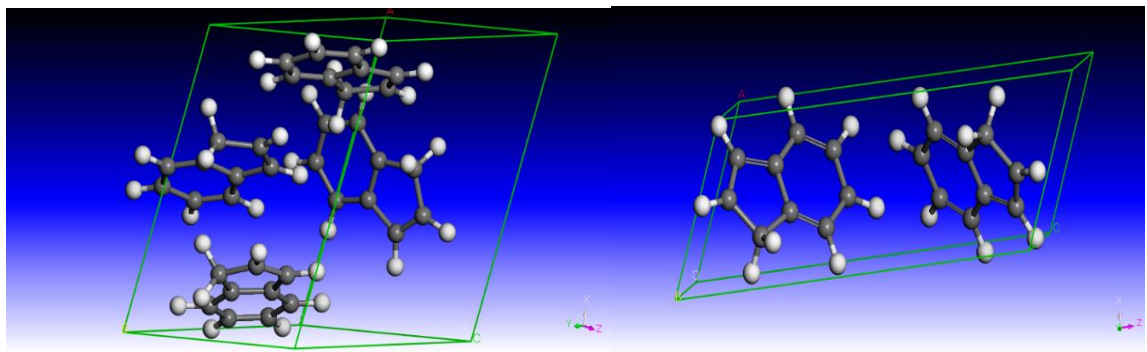

Figure 3. Optimized structures of amorphous unit cell (left) and crystalline unit cell (right) of indene.

The calculated IR spectra of amorphous and crystalline indene are compared with the experimental spectra in Figure 4. A general good agreement is found, the calculated spectra reproducing the main peaks positions and intensities. The simulations reproduce fairly well the relative intensities between the different absorptions. Although some frequency deviations are observed, for the main vibrations it is possible to make a clear correlation between the theoretical and experimental bands. For example, for the crystalline solid, the strongest absorption, L9, is predicted only 11 cm$^{-1}$ shifted to lower wavenumber, and L5 and L4, only 25 and 50 cm$^{-1}$ shifted to higher wavenumbers, respectively. The larger frequency shifts (approximately 5%) are in the CH stretching modes, that are predicted at about 120 cm$^{-1}$ higher wavenumbers. For the amorphous solid, frequency deviations show a similar tendency. The theoretical calculations provide information on the atomic displacements associated to each particular vibration mode, that can be visualized using CASTEP. Indene solids are molecular solids where the indene molecules are bonded via hydrogen bonds, and it is possible, to some extent, to associate the vibrations of the solid (phonons) to specific vibrations of the indene molecule. However, the mode assignment is not always clear. In some cases, the vibrations can by clearly attributed to a particular mode, but in others there is mode mixing and the assignment is vaguer. With that limitation, an assignment of the strongest bands of indene is presented in Table 1, which is in good agreement with previous assignments of the vibrational spectra of indene in the gas and liquid phase (Klots 1995).

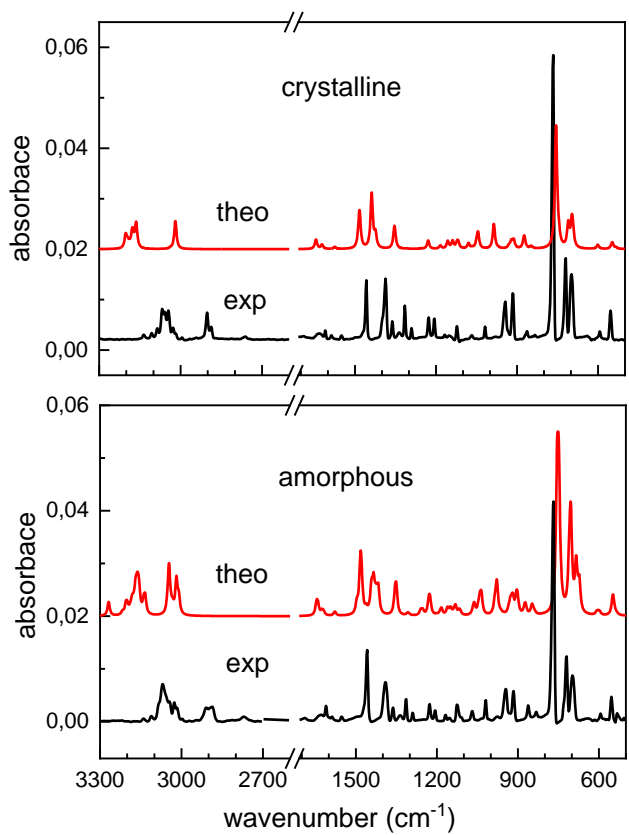

Figure 4. Comparison of theoretical (red) and experimental (black) spectra of solid indene. Top panel: crystalline indene. Bottom panel: amorphous indene. In the theoretical spectra the calculated mode intensities are represented with 10 cm$^{-1}$ FWHM gaussians and scaled in intensity for better comparison with the experimental spectra. The theoretical spectra have been offset in the vertical axis for better visualization.

The theoretical calculations provide also absolute infrared band strengths, that have been listed in Table 5 for the amorphous and crystalline phases. In the last two columns, the deviations between theoretical and experimental band strengths are listed. For the two forms the agreement between experimental and theoretical values is good, given the approximations implicit in the derivation of both experimental and theoretical magnitudes. It can be seen that for some absorptions the agreement is within 10%, whereas for others it is in the 50-70 % range. For example, the intensities of the strongest absorptions, L1 ($\nu_a$ CH) and L9 ($\delta$ CH in-phase out-of-plane) are very well reproduced both for amorphous and crystalline solids. On the other hand, the intensity of the L2 ($\nu_a$ CH2) mode is poorly reproduced in both models. On average, for the eleven bands considered, the disagreement is below 30%, being better for the crystalline form. Regarding the intensity changes occurring upon crystallization that are observed experimentally (see last column in Table 2), the calculations are not able to reproduce them, and therefore do not allow to provide a chemical explanation of the effect. Nonetheless, they do predict the intensity increase from 2.0 to 2.17 x10$^{-17}$ cm/molec of the strong L9 ($\delta$ CH in-phase out-of-plane) mode upon crystallization, a tendency that is in agreement with the experimental observations.

Table 5. Calculated peak positions and band strengths of the strongest IR bands of amorphous and crystalline indene. The intensity of the modes within the interval indicated in columns two and five has been added to estimate the bands strengths given. $\Delta\nu_{exp\text{-}theo}$ is the deviation between theoretical and experimental peak positions. The ratio between experimental and theoretical band strengths in given in the last two columns.

| Band label | Peak amorph. (cm$^{-1}$) | Interval amorph. (cm$^{-1}$) | $\Delta\nu_{exp\text{-}theo}$ amorph. (cm$^{-1}$) | A' amorph. (10$^{-17}$ cm/molec) | Mode Interval cryst. (cm$^{-1}$) | Peak position cryst. (cm$^{-1}$) | $\Delta\nu_{exp\text{-}theo}$ Cryst. (cm$^{-1}$) | A' cryst. (10$^{-17}$ cm/molec) | A$_{exp}$/A$_{theo}$ amorph. | A$_{exp}$/A$_{theo}$ cryst. |
|---|---|---|---|---|---|---|---|---|---|---|
| L1 | 3161 | 3268-3133 | -92 | 1,14 | 3202-3159 | 3177 | -120 | 1,07 | 1,19 | 0,98 |
| L2 | 3031 | 3050-3009 | -133 | 0,88 | 3038-3019 | 3018 | -116 | 0,47 | 0,48 | 0,66 |
| L3 | 1632 | 1647-1618 | -10 | 0,23 | 1644-1621 | 1631 | -11 | 0,23 | 0,65 | 0,61 |
| L4 | 1482 | 1496-1460 | -23 | 0,70 | 1487-1482 | 1483 | -25 | 0,67 | 0,91 | 0,72 |
| L5 | 1434 | 1443-1416 | -44 | 0,77 | 1438-1424 | 1438 | -50 | 1,12 | 0,66 | 0,61 |
| L6 | 1352 | 1356-1349 | -38 | 0,33 | 1355-1353 | 1353 | -38 | 0,39 | 0,42 | 0,56 |
| L7 | 979 | 995-952 | -34 | 0,37 | 1049-1044 | 920 | 24 | 0,32 | 1,03 | 1,44 |
| L8 | 912 | 950-899 | 4 | 0,49 | 986-985 | 873 | 42 | 0,41 | 0,53 | 0,95 |
| L9 | 751 | 758-743 | 17 | 2,01 | 756-745 | 755 | 11 | 2,14 | 1,00 | 1,07 |
| L10 | 705 | 710-703 | 15 | 0,45 | 709-711 | 710 | 10 | 0,40 | 1,33 | 1,78 |
| L11 | 678 | 690-672 | 18 | 0,64 | 697-692 | 696 | 2 | 0,59 | 0,95 | 1,31 |
| L12 | 547 | 550-540 | 6 | 0,19 | 548-533 | 547 | 6 | 0,13 | 0,95 | 1,92 |

## 4.4. Indene in water ice.

Ice mixtures with high dilutions, 7% to 2% number molecules of indene in water, were generated by simultaneous deposition of both gases at 10 K, with the goal of studying the possible spectral changes in the indene spectra caused by the water ice matrix. Figure 5 shows the IR spectra of the mixtures together with a pure indene spectrum. The stoichiometry has been estimated from the intensity of a water band and an indene band in the IR spectrum of the mixture. The OH stretching for water and the in-phase out-of-plane CH bending (L9) for indene, with band strengths of $1.9\ 10^{-16}$ cm/molec (Mastrapa et al. 2009) and $2.0\ 10^{-17}$ cm/molec (Table 2 this work), respectively, were chosen. Possible variations in the band strengths due to the mixture have been neglected. In the present work the band strengths of the mixtures are not provided. We refer to the recent works by Hudson and coworkers (Hudson et al. 2022, Gerakines et al. 2022) where they measured band strengths variations in water ice environments of several molecules, both polar (HCN) and non-polar ($C_6H_6$) and found them to be below 10%.

In general, the IR spectrum of indene is not much affected by the water ice matrix, consistent with previous larger PAH studies (Bernstein et al. 2005). In contrast with other astrophysical complex organic molecules (COMs) like urea (Timón et al. 2021) or glycine (Maté et al. 2011), previously investigated in our group, the IR spectrum of indene ice presents narrow lines, with FWHM of the order of 10 cm$^{-1}$. This characteristic makes the sharp indene bands easily distinguishable on top of the broad ASW absorption profiles, at least in the 1500-500 cm$^{-1}$ region, away from the strong OH stretching mode of water (see Figure 5). In the figure inserts it is shown how the band profiles of the L9, L10, L11 bands broaden in the spectrum of the mixture, making these indene features less visible than L4 or L5. The latter two band are not very much affected by the presence of the ice matrix, and do not broaden appreciably. It seems that the water ice hydrogen bond network has an effect on the low wavenumber (low energy) indene vibrations assigned to in-phase out-of-plane CH bending (L9, L10, L11); and this effect is not so strong for higher wavenumber vibrations, in particular to those assigned also to an out-of-plane but out-of-phase (L4) CH bending, and to the CH2 bending (L5).

Looking at the high wavenumber region, the CH stretching bands L1 and L2 do not suffer an appreciable broadening and the only effect that the water ice matrix has is modifying the relative intensities of the features within the L1 or L2 profiles. However, these intensity variations could be affected by baseline subtraction, which is difficult to do in this region because it coincides with the tail of the strong OH absorption.

Regarding wavenumber shifts, the effect of the water ice matrix is very small, nonetheless, the peak positions of the low wavenumber peaks of indene in the 2% mixture are listed in Table 6, together with those of the pure species. A similar behavior had been previously observed for naphthalene in water ice (Sandford et al. 2004).

As for ASW, the only modification appreciated in the spectra of the mixtures appears in the 3700 cm$^{-1}$ dangling bond (DB) region, the DB feature being modified by the presence of impurities (Michoulier et al 2020).

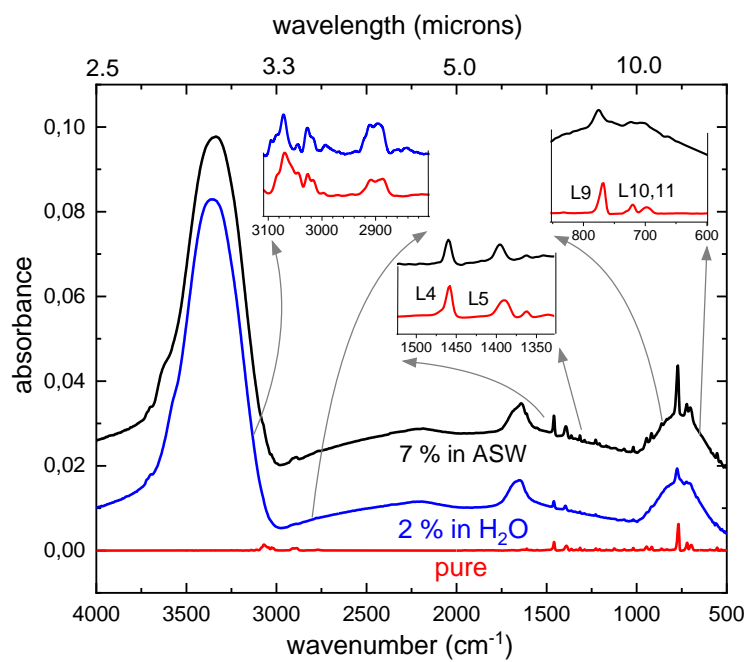

Figure 5. IR spectra of 7% (blue) and 2 % (black) indene in ASW at 10 K compared with a pure indene ice at the same temperature (red). Baseline corrections and vertical offsets have been applied to the spectra for a better visualization.

Table 6. Peak positions of some indene bands in a 2% mixture in amorphous solid water (ASW) at 10 K, compared to those of the pure species amorphous spectrum.

| label | Peak position | | | |
|---|---|---|---|---|
| | pure | | 2% in ASW | |
| | cm$^{-1}$ | μm | cm$^{-1}$ | μm |
| L4 | 1459 | 6,85 | 1461 | 6,84 |
| L5 | 1390 | 7,19 | 1395 | 7,17 |
| L6 | 1314 | 7,61 | 1314 | 7,61 |
| L7 | 945 | 10,58 | 945 | 10,58 |
| L8 | 916 | 10,92 | 920 | 10,87 |
| L9 | 768 | 13,02 | 774 | 12,92 |
| L10 | 720 | 13,89 | 722 | 13,85 |
| L11 | 698 | 14,33 | 702 | 14,25 |

## 5. Astrophysical implications

At the very low temperature (10 K) of TMC-1, volatiles should be largely frozen in the ice mantles of dust grains. Following Dartois et al. (2022) we express the partition of indene between the gas and the solid as:

$$f_{H,gas}(indene) = C \, \chi_{ice}(indene) \qquad (3)$$

where $f_{H, gas}$ is the abundance of indene gas-phase molecules with respect to the number of H atoms (which for a cold cloud will be mostly in the form of $H_2$), $\chi_{ice}(indene)$ represents the fraction of indene molecules in the ice, and C is a factor summarizing the effects of the processes leading to the increase and depletion of indene in each phase. For the conditions of a cold cloud like TMC-1, Dartois et al. (2022) consider that sputtering of small PAHs -containing ice by cosmic rays is the source of that gas-phase PAHs, and that the molecule is depleted from the gas-phase either by VUV photolysis or by condensation on the solid grains. Using experimental sputtering yields, Dartois et al. (2022) estimated a value C = 1.9 - 3.2 x $10^{-7}$ for naphthalene highly diluted in an ice matrix, and similar values are expected in general for small PAHs. By analogy, we assume C = 2 - 3 x $10^{-7}$ for indene.

Using equation (3), we can express the amount of indene in the ice as a function of the amount of water:

$$N_{ice}(indene) \approx \frac{f_{H,gas}(indene)}{C} \frac{N_{ice}(H_2O)}{\chi_{ice}(H_2O)} \qquad (4)$$

where $N_{ice}(indene)$ and $N_{ice}(H_2O)$ are the ice column densities of indene and water respectively and $\chi_{ice}(H_2O)$ is the fraction of water molecules in the ice. The ice mantles in the dust grains of TMC-1 are observed in the IR extinction spectra toward the background star Elias 16 (Whittet et al. 1988, Smith et al. 1993, Chiar et al. 1996). From these observations, the fraction of water in the ice is estimated at $\chi_{ice}(H_2O) \approx 0.65$, and the water-ice column density at $2.5 \times 10^{18}$ cm$^{-2}$ (Knez et al. 2005). Substituting these values in equation (4) and taking the gas-phase indene abundances reported by Cernicharo et al. (2021) and Burkhardt et al. (2021), the column density of indene in the ice mantles derived from equation (4) is $N_{ice}(indene)=1.5-0.6 \times 10^{16}$ cm$^{-2}$ and the corresponding fraction of indene molecules in the ice is $\chi_{ice}(indene) = (6 - 2.4) \times 10^{-3}$. The relevant column densities are summarized in Table 7. According to these estimates, the fraction of gas phase vs solid indene is $\approx 10^{-3}$.

Table 7. Column densities in TMC-1

| | Column densities TMC-1, Elias 16 | | | |
|---|---|---|---|---|
| | H[a] | $H_2O$[b] | Indene (gas)[a] | Indene (ice)[c] |
| N (cm$^{-2}$) | $2 \times 10^{22}$ | $2.5 \times 10^{18}$ | $(1.6 - 1) \times 10^{13}$ | $(1.5 - 0.6) \times 10^{16}$ |

a) Cernicharo et al. (2021), Burkhardt et al. (2021)
b) Knez et al. 2005
c) This work, equation (4) with C= 2-3 × $10^{-7}$. See the text

As mentioned in the introduction, tentative signatures of PAHs have been suggested in the IR extinction spectra of ices and some of them have been even used for the estimation of the amount of cosmic carbon locked in PAHs in dense clouds (Bowman et al. 2011, Hardegree-Ullman et al. 2014, Chiar et al. 2021 ). The preferred band for this type of estimates is the CH stretch vibration appearing at about 3.3 µm because its position is very stable for different PAHs and is thus adequate for the mixtures of PAHs presumed to be present in the dust grains. The 3.3 µm feature is not the most intense IR band for individual PAHs, which usually corresponds to the CH out-of-plane bending vibrations appearing at wavelengths larger than 11 µm, but the CH out-of-plane bending bands are more variable in position than the CH stretch and lead to broadened and weaker bands in the spectra of PAH mixtures (Allamandola et al. 1999) that can´t be easily discerned from other ice components.

Two IR absorption features at 3.3 and 3.47 µm, have been reported in the line of sight toward Elias 16 which is close to the position of TMC-1 (Chiar et al. 1996, Chiar et al. 2021). They can be seen in the upper curve of Figure 6. Note that these bands have been derived from ground-based observations of a very weak substructure in the long wavelength wing of the OH band of water ice and may be affected by a large uncertainty due to baseline subtraction and telluric absorptions. The full absorption spectrum of the OH band of ice in Elias 16, including data from the ISO satellite, can be found in Whittet et al. (1998) and Gibb et al. (2004). The 3.3 µm feature, is generally attributed to aromatic CH stretching vibrations (Sellgren et al. 1994, Sellgren 1995, Bowman et al. 2011, Hardegree et al. 2014, Chiar et al. 2021). The 3.47 µm absorption is mostly assigned to ammonia hydrates (Dartois & D' Hendecourt 2001, Dartois et al. 2002, Boogert el al. 2015) but other carriers have been also suggested, including H atoms bound to ternary $sp^3$ carbon (Allamandola et al. 1992), and hydrogenated PAHs (Chiar et al. 2021). Note also that in some of the mentioned observations a contribution from crystalline ice could lead to distortions in the baseline subtraction. Even if ammonia hydrates were the predominant carriers, contributions from other species cannot be excluded. With the value of $N_{ice}$(indene) from Table 7 and the experimental band strengths from Table 2 we can now estimate the expected contribution of indene to these absorption features.

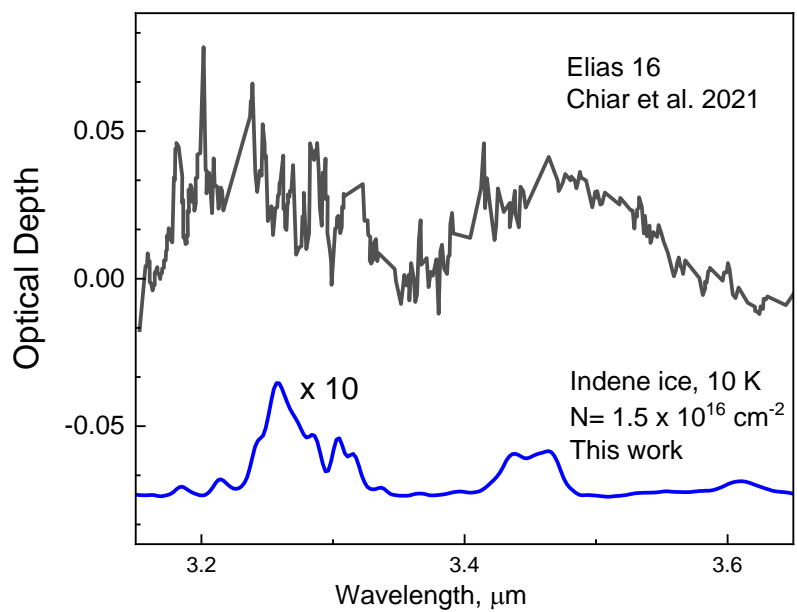

Figure 6. Upper curve: Observed absorption spectrum towards Elias 16 (from Chiar et al. 2021). For the whole spectrum of the OH band see: Whittet et al. (1998), Gibb et al. (2005). Lower curve: Absorption spectrum of indene deposited at 10 K (Fig. 1) scaled for a column density $N_{ice}$(indene)= $1.5 \times 10^{16}$ cm$^{-2}$ with the band strengths of Table 2. The selected column density corresponds to the higher value estimated above for indene in the ice mantles (Table 7). The intensity of this spectrum has been multiplied by 10 in the figure.

Indene has also two absorption bands (L1 and L2) in this spectral region. An optical depth spectrum of indene ice at 10 K is also shown in Figure 6 for comparison. It has been scaled for a column density $N_{ice}$ (indene) = $1.5 \times 10^{16}$ cm$^{-2}$, which corresponds to the higher amount of indene estimated above (Table 7) for the ice mantles of TMC-1 (using the lower estimate would reduce the intensity by a 2.5 factor). For a better appreciation, the indene optical depth spectrum is multiplied by 10 in the figure. The integrated area of the L1 band (aromatic CH stretch at 3.25 μm) of indene in the optical depth spectrum of Figure 6 is roughly 0.2 cm$^{-1}$ and corresponds to about 5 % of the area of the first broad band in the observational spectrum. Comparable contributions to this absorption band can be expected from benzene and naphthalene, whose cyano derivatives have also been observed in TMC-1 (Sita et al. 2022), which strongly suggests that an appreciable fraction of the 3.3 μm absorption feature toward Elias 16 could be accounted for by indene, benzene, naphtalene and other small aromatic hydrocarbons formed in situ in TMC-1, and not just by a mixture of large PAHs from post AGB stars, with an average of 50 carbons, as has been traditionally assumed. The L2 band of indene (CH stretch of the CH$_2$ group at 3.45 μm) in the spectrum of Fig. 6 has an area of approximately 0.07 cm$^{-1}$, which represents ≈ 2 % of the second observational band and supports the assignment of this feature, at least in a small part, to hydrogenated (not necessarily large) PAHs (Chiar et al. 2021). For the same column density of indene, other characteristic absorption lines like L3 (6.18 μm), L4 (6.86 µm) and L9 (13,02 µm), which are outside the range of Figure 6, have peaks, of ≈ 7 × $10^{-4}$, ≈ 4 × $10^{-3}$ and ≈ 1 × $10^{-2}$ respectively in the optical depth spectrum, and would be buried under broad ice absorptions (Chiar et al. 2021, Knez et al. 2005) due to multiple components (Boogert et al. 2015)

Finally, assuming a cosmic carbon abundance of 3.24 × $10^{-4}$ with respect to H (Hensley and Draine, 2021) and taking into account that indene molecules have 9 C atoms, the fraction of cosmic carbon locked up in indene in TMC-1, mostly in ice mantles, would be ≈ (2 - 0.8) × $10^{-2}$. This value is consistent with the carbon inventory, but is possibly too high for a single small hydrocarbon. Further work is needed to assess the presence and stability of indene and other small aromatic hydrocarbons in the ice mantles of dust grains in cold clouds.

## 6. Summary and conclusions.

Two solid phases of indene, amorphous and polycrystalline have been generated via vapor deposition on a cold surface. An amorphous form is formed in the 10 K deposit, and a crystalline form for deposition at 160 K. The phase transition from amorphous to crystalline is observed to take place between 120 K and 130 K when worming the ice at 1 K/min. The solid sublimates at 180 K under high vacuum conditions. Infrared spectra and infrared band strengths are provided for both amorphous and crystalline phases. The infrared spectra of highly diluted mixtures (2% and 7%) of indene in amorphous solid water generated at 10 K are also provided. The indene spectrum is not much altered by the water ice environment. Small frequency shifts, not larger than 6 $cm^{-1}$ are appreciated, and only the bands with wavelengths below 1000 $cm^{-1}$ are significantly broadened.

Crystalline and amorphous solids of indene have been constructed using density functional theory, and their infrared spectra have been calculated. Comparison of the calculated spectra with the experimental one, strongly suggests that the most probable crystalline structure of indene has monoclinic symmetry. No previous theoretical nor experimental information about the crystalline structure of indene was available in the literature. The density found for the crystal is 1.40 g/cm3, larger than the experimental value known for the liquid (1.3 g/$cm^3$). For the amorphous form, the most stable structure was found to have a density of 1.27 g/cm3, close to that of the liquid. Assignments of the main absorptions of solid indene, based on the calculations, are given. Theoretical IR band strengths are also given and compare well with the experimental ones, within experimental error.

Our results are expected to help the search of this species in the solid phase in astrophysical environments with the JWST. They could be applied also in laboratory astrochemistry. The band strengths will be an important magnitude to estimate the number of indene molecules in the ice deposit, for example, when conducting energetic processing experiments. With the signal to noise ratio of our experimental setup, 0.015 (mean square deviation of the baseline noise in the transmittance spectrum), indene could be detected in 1% mixtures using the 1461 $cm^{-1}$ (7,14 µm) (L3) band, and in 2% mixtures when looking at the L8 band (768 $cm^{-1}$, 13,00 µm). Larger fractions, about 7%, are needed if looking at L1, L2 in the 3000 $cm^{-1}$ (3,3 µm) region.

Using the observed indene gas-phase column density from mm observations of rotational transitions, and assuming that indene molecules arise from the cosmic ray ejection from the ice mantles, we estimate with a simple literature model a column density of $N_{ice}$ (indene) = (1.5 -0.6) × $10^{16}$ cm$^{-2}$ for the ice mantles of TMC-1. With our measured band strengths, this amount of solid indene could account for 2-5 % of the intensity of the weak absorption feature at 3.3 μm in the IR spectrum towards Elias 16 reported by Chiar et al. (2021). This suggests that small polyaromatic hydrocarbons formed in situ through cloud chemistry, and not just large PAHs from post AGB stars, could contribute appreciably to hypothetical PAH signatures in the IR ice spectra. Assuming a cosmic carbon abundance of 3.24 × $10^{-4}$ with respect to H (Hensley and Draine, 2021), the fraction of cosmic carbon locked up in indene in TMC-1, mostly in ice mantles, would be ≈ (2 - 0.8) × $10^{-2}$. This value is consistent with the carbon inventory, but is possibly too high for a single small hydrocarbon. The occurrence, stability and chemical relevance of small aromatic hydrocarbons in the ice mantles should be further investigated and considered in astrochemical models.

**ACKNOWLEDGEMENTS**

This work was funded by Ministerio de Ciencia e Innovación (MCI) of Spain under grant PID2020-113084GB-I00, and by the European Union under grant ERC-2013-Syg-210656- NANOCOSMOS. The computation time provided by the Centro Técnico de Informática, cluster Trueno from CSIC and Centro de Supercomputación de Galicia CESGA is deeply acknowledged.

**Data Availability Statements**

Data available on request. The data underlying this article will be shared on reasonable request to the corresponding author.

**References.**

Allamandola, L.J., Tielens, A.G.G.M., and Barker, J.R. 1985, ApJ 290, L25. doi:10.1086/184435
Allamandola, L. J., Hudgins, D. M. and Sandford, S. A. 1999, ApJ, 511, L115
Allamandola, L. J., Sandford, S. A., Tielens, et al., 1992, ApJ, 399, 134

Baroni, S., de Gironcoli, S., Dal Corso,S., Giannozzi, P., 2006, Reviews of Modern Physics, 73, 515
Bowman, J., Mattioda, A. L., Linnartz, H. and Allamandola, L. J. 2011, A&A, 525, A93
Bregman, J. D., Hayward, T. L., and Sloan, G. C., 2000, ApJ, 544, L75
Bernstein, M. P., Sandford, S. A., Allamandola, L. J., 2005, ApJSS, 161:53
Boogert, A.C.A., Gerakines, P. A., and Whittet, D.C.B., 2015, Annu. Rev. Astron. Astrophys. 56, 541
Brooke, T. Y., Sellgren, K., and Geballe, T. R. 1999, 517, 883
Burkhardt, A.M., Lee, K.L.K., Changala, B, et al. 2021, ApJ. 913, L18. doi:10.3847/2041-8213/abfd3a
Caminati, W. 1993, J. Chem. Soc. Faraday Trans., 89, 4153
Cernicharo, J. 2004, ApJ, 608, L41
Cernicharo, J., Heras, A. M., Tielens, A. G. G. M. et al. 2001, ApJL, 546, L123

Cernicharo, J., Agúndez, M., Cabezas, C., et al. 2021, A&A, 649, L15. doi:10.1051/0004-6361/202141156

Cernicharo, J., Agúndez, M., Kaiser, R. I., Cabezas, C., et al. 2021, A&A, 652, L9

Cernicharo J., Agúndez, M., Cabezas, C., Tercero, B., et al. 2021, A&A 649, L15

Cherchneff, I. 2012. A&A, 545, A12. doi:10.1051/0004-6361/201118542

Chiar, J., Adamson, A. J., and Whittet, D. C. B., 1996, ApJ, 472, 265

Chiar, J. E., de Barros, A. L. F., Mattioda, A. L., and Ricca, A. 2021, ApJ, 908, 239

Clark,S. J., Segall, M.D., Pickard, C.J., Hasnip, P., Probert, M.I.J., Refson, K.,Payne, M.C., 2005, Zeitschrift für Kristallographie , 220, 567

Curnow,O.J., Tsuruta, K., 2012, Zeitschrift für Kristallographie - New Crystal Structures, 227,127

Dartois, E., & d'Hendecourt, L. 2001, A&A, 365, 144

Dartois, E., & d'Hendecourt, L., Thi, W., at al. 2002, A&A, 394, 1057

Dartois, E., Chabot, M., Koch, F. et al. 2022, A&A, 663, A25

Doddipatla, S., Galimova, G. R., Wei, H., et al. 2021, Sci. Adv. 7, eabd4044

El-Azhary, A. A. 1999, Spectr. Acta A, 55, 2437

Frenklach, M., and Feigelson, E. 1989, ApJ 341, 372. doi:10.1086/167501

Gerakines, P. A, Yarnall, Y.Y., Hudson, R.L., 2022, MNRAS 509, 3515

Gibb, E. L., Whittet, D. C. B., Boogert, A. C. A., et al. 2004, ApJSS, 151, 35

Gupta, D., Choi,H., Singh,S., Modak,P., Antony,B., Kwon,D-C, Song, M-Y, Yoon, J-Y, 2019, J. Chem. Phys, 150, 064313

Goesmann F. et al., 2015, Science, 349, aab0689

Hamdouni, N., Boudjada,A., Medjroubi, M.L., 2019, Journal of Molecular Structure, 1198 126827

Hardegree-Ullman, E. E., Gudipati, M. S., Boogert, A. C. A. et al. 2014, ApJ, 784, 172

Hensley, B. S., and Draine, B. T. 2021, ApJ, 906, 73

Hudson, R.L., Yarnall, Y.Y., 2022, Icarus, 377, 114899

Keane, J. V., Tielens, A. G. G. M., Boogert, A.C. A., Shutte, W. A., and Whittet, D. C. B. 2001, A&A, 376, 254

Klots, T. D. 1995, Spectr. Acta A, 51, 2307

Knez, C., Boogert, A. C. A., Pontoppidan, K. M. et al. 2005, ApJ, 635, L145

Kwok, S., 2004, Nature, 430, 985

Léger, A., and Puget, J.L. 1984, A&A, 137, L5

Malek, S. E., Cami, J. and Bernard-Salas, J. 2012, 744, 16

Martínez, L., Santoro, G., Merino, P. et al. 2020, Nat. Astron. 4, 97

Marín-Doménech, R., et al., 2015, A&A 584, A14, DOI: 10.1051/0004-6361/201526003

Mastrapa R. M., Sandford S. A., Roush T. L., Cruikshank D. P., Dalle Ore C. M., 2009, ApJ, 701, 1347

Maté B., Medialdea, A., Moreno, M. A., Escribano,R., and Herrero, V. J., 2003, J. Phys. Chem. B, 107, 11098-11108

Maté B., Rodríguez-Lazcano Y., Galvez O., Tanarro I., Escribano R., 2011, Phys. Chem. Chem. Phys., 13, 12268

Maté et al.,2017, MNRAS 470, 4222, doi:10.1093/mnras/stx1461

McGuire, B. A., Burkhardt, A. M., Kalenskii, S. V., et al. 2018, Science, 359, 202
McGuire, B. A., Loomis, R. A., Burkhardt, A. M. et al. 2021, Science, 371, 1265
Michoulier, E.,Toubin, C., Simon, A., Mascetti, J., Aupetit, C., and Noble, J. A., 2020, J. Phys. Chem. C, 124, 2994
Payne, M. C.; Teter, M. P.; Ailan, D. C.; Arias, A.; Joannopoulos, J. D., 1992, Rev. Mod. Phys., 64, 1045
Pickard, C.J., Winkler, B., Chen, R.K, Lee,M.H., Lin, J.S., White, J.A., Milman, V., Vanderbilt,D.,2000, Physical Review Letters, 2000, **85,** 5122-5125. DOI: 10.1103/PhysRevLett.85.5122

Sandford, S. A., Bernstein, M, P. and Allamandola L. J. 2004, ApJ, 607, 346

Santoro, G., Martínez, L., Lauwaet, K. et al. 2020, ApJ, 895, 97
Sellgren, K., Smith, R. G. and Brooke, T. Y. 1994, ApJ, 433, 179
Sellgren, K., Brooke, T. Y., Smith, R. G., and Geballe T. R. 1995, ApJ, 449, L69
Smith, R. G., Sellgren, K., and Brooke, T. Y. 1993, MNRAS, 263, 749
Sita, M. L., Changala, P. B., Xue, Ci., et al. 2022, ApJL, 938, L12
Tkatchenko, A., Scheffler, M., 2009, Physical Review Letters, 102, 073005
Tielens, A.G.G.M. 2008, Annu. Rev. Astron. Astrophys. 46, 289
Timón V., Maté B., Herrero V. J., Tanarro I., 2021, Phys. Chem. Chem. Phys., 23, 22344

Whittet, D. C. B., Bode, M. F., Longmore, A. J., et al. 1988, MNRAS, 233, 321
Whittet, D. C. B., Gerakines, P. A.,Tielens, A.G.G.M., et al. 1998, ApJ 498, L159
Willner, S.P., Puetter, R.C., Russell, R.W., at al. 1979, Astrophys. Space Sci. 65, 95, doi:10.1007/BF00643492
Woods, P. M., Millar, T. J., Herbst, E. et al. 2003, A&A, 402, 189
Zhang, Y., and Kwok, S. 2015, ApJ 798, 37. doi:10.1088/0004-637X/798/1/37
Zhang,Y., Yang,W., 1998, Physical Review Letters, 80, 890

**APPENDIX A1. CALIBRATION OF THE QUADRUPOLE MASS SPECTROMETER**

For the calibration of the QMS we used the infrared spectra of water ice. We measured the intense OH absorption band between 4939 and 2969 cm$^{-1}$ of water ice deposited at 10 K. The column density, $N_{H2O}$, of water molecules in the ice is given by

$$N_{H2O} = 2.303 \frac{Int}{A'_{H2O}} \qquad (A1)$$

where $A'_{H2O}$ is the band strength of the OH band and *Int* is the band area of the OH band in the absorbance spectra. If we assume a sticking probability of one for water molecules hitting the substrate at 10 K, the column density can be expressed as:

$$N_{H2O} = n_{H2O} \left(\frac{kT}{2\pi m}\right)^{1/2} \Delta t \qquad (A2)$$

where $n_{H2O}$ is the molecular density, held constant during deposition, $T$ is the gas temperature, $k$, the Boltzmann constant, and $m$ the molecular mass. The molecular density of $H_2O$ is readily derived as:

$$n_{H2O} = 2.303 \frac{Int}{A'_{OH} \Delta t} \left(\frac{2\pi m}{kT}\right)^{1/2} \qquad (A3)$$

The integral can be obtained from the measured integrated absorbance spectra of water ice over the frequency range of interest and the band strength is taken from the literature $A'_{OH} = 1.9 \times 10^{-16}$ cm molec$^{-1}$ (Mastrapa et al. 2009). During the water ice deposition process, the QMS signal corresponding to mass 18 can be expressed as: $I_{18} = c \times n_{H2O}$, where c is a proportionality constant that can be determined once $n_{H2O}$ is derived from equation (A3).

For the calibration of the QMS for indene, we used $H_2O$ as reference. The different detection efficiencies for the two species were corrected by considering the dependence of the QMS signals, $I_i$, on ionization cross section and fragmentation pattern (Martín-Doménech et al. 2015):

$$I_i \propto n_m \sigma_m f_i \qquad (A5)$$

where $I_i$ is the QMS reading for peak i from species m, $n_m$ the molecular density of species m, $\sigma_m$ its electron impact ionization cross section, and $f_i$ is the fraction of the total m signal corresponding to peak i. The indene molecular density in the deposition chamber, taken the mass 116 in the QMS for the indene molecule, is given by:

$$n_{C9H8} = \frac{I_{116} \times n_{H2O} \times \sigma_{H2O} \times f_{18}}{I_{18} \times \sigma_{C9H8} \times f_{116}} \qquad (A6)$$

Equation (A6) assumes tacitly that the detection sensitivity is roughly the same for the two mass fragments, which can be taken as a good approximation for the 200 uma QMS used. The cross section for water, $\sigma_{H2O} = 2.53$ Å$^2$, was taken from table S4 of Goesmann et al. (2015) and it was calculated for an electron energy of 70 eV (as employed in the QMS). No experimental neither theoretical value was found in the literature for the ionization cross section of indene. However, a recent paper by Gupta et al. (2019) provides calculated values for several cyclic organic molecules, in particular double-ring species like benzothiophene and naphthalene. Indene ionization cross section (for 70 eV electron energy) was assumed to be the same as that of naphthalene, $\sigma_{C9H8} = 22$ Å$^2$. From the respective QMS electron impact fragmentation patterns of water and indene, $f_{18} = 0.81$ and $f_{116} = 0.5$ were derived. Using these values, $n_{C9H8}$ can be derived, and then the calibration constant c' for indene as: $I_{116} = c' \times n_{C9H8}$.

There are several sources of error associated with the calibration procedure just described (assumed sticking coefficient, pressure fluctuations during deposition, ionization cross section values, different transmission efficiency of the two masses through the QMS,...), which are difficult to assess precisely. We assume a conservative 30% uncertainty for the $n_{C9H8}$ values derived with this procedure.